\documentclass{llncs}

\newcommand{\iffull}[1]{#1}
\newcommand{\ifconf}[1]{}

\usepackage{llncsdoc}

\usepackage[utf8]{inputenc}
\usepackage[T1]{fontenc}
\usepackage{amssymb}
\usepackage{multicol}
\usepackage{hyperref}
\usepackage{listings}
\usepackage{mathtools}
\usepackage{algpseudocode}
\usepackage{semantic}
\usepackage{url}
\usepackage{paralist}
\usepackage{pgfplots}
\usepackage{tabularx}
\usepackage{hhline}
\usepackage{caption}
\usepackage[absolute]{textpos}

\begin{document}
\iffull{
\setlength{\TPHorizModule}{\linewidth}
\setlength{\TPVertModule}{\TPHorizModule}
% \textblockorigin{8.5mm}{3mm} 
\textblockorigin{42mm}{10mm} 
\begin{textblock}{1}(0,0)
  \fbox{\parbox{\textwidth}{\scriptsize\copyright Springer-Verlag,
      2017.  This is the author's version of the work. It is posted
      here by permission of Springer-Verlag for  
      your personal use. Not for redistribution. The final 
      publication is published in the proceedings of the 22nd
      European Symposium on Research in Computer Security (ESORICS  '17),
      and is available at 
      \href{http://www.link.springer.com}{link.springer.com} 
%   (\href{http://dx.doi.org/10.1007/978-3-642-11747-3_4}{DOI})
}}
\end{textblock}
}
% Copyright
% \setcopyright{acmcopyright}

\newcommand{\sys}{\mbox{WebPol}}

\newcommand{\TODO}[1]{\textcolor{red}{#1}}
\newcommand{\CH}[1]{\textcolor{red}{CH: #1}}
\newcommand{\DG}[1]{\textcolor{red}{DG: #1}}
\newcommand{\VR}[1]{\textcolor{red}{VR: #1}}

\newtheorem{mydef}{Definition}
\newtheorem{myLemma}{Lemma}
\newtheorem{myThm}{Theorem}
\newtheorem{myaxiom}{Axiom}
\newtheorem{myAssumption}{Assumption}

\newcommand {\mV} {\mathcal{V}}
\newcommand {\mL} {\mathcal{L}}
\newcommand {\mH} {\mathcal{H}}
\newcommand {\mlp} {\mathsf{low}}
\newcommand{\mC}{\mathcal{C}}
\newcommand {\PC} {\mathit{pc}}

\newcommand {\Succ} {\mathit{Succ}}
\newcommand {\Left} {\mathit{Left}}
\newcommand {\Right} {\mathit{Right}}
\newcommand {\src} {\mathit{src}}
\newcommand {\dst} {\mathit{dst}}
\newcommand {\isIPD} {\mathit{isIPD}}
\newcommand {\IPD} {\mathit{IPD}}
\newcommand {\ipd} {\mathit{ipd}}
\newcommand {\cf} {\mathit{CF}}
\newcommand {\cond} {\mathit{cond}}
\newcommand {\push} {\mathit{push}}
\newcommand {\false} {\mathit{false}}
\newcommand {\true} {\mathit{true}}
\newcommand {\val} {\mathit{value}}
\newcommand {\base} {\mathit{base}}
\newcommand {\Empty} {\mathit{empty}}
\newcommand {\eV} {\mathit{excValue}}
\newcommand {\SC} {\mathit{sc}}
\newcommand {\LO} {\Lambda}
\newcommand{\CFG}{\mathit{CFG}}
\newcommand{\phase}{\mathcal{P}}
\newcommand{\SP}{\mathcal{S}}
\newcommand{\CP}{\mathcal{C}}
\newcommand{\TP}{\mathcal{T}}
\newcommand{\BP}{\mathcal{B}}
\newcommand{\DP}{\mathcal{D}}
\newcommand{\aL}{\mathcal{A}}
\newcommand{\nL}{\mathcal{N}}
\newcommand{\eHL}{\mathcal{H}}
\newcommand{\pl}{\mathit{\ell_p}}
\newcommand{\dE}{\mathit{dispatchEvent}}
\newcommand{\gEP}{\mathit{getEventPath}}
\newcommand{\gEH}{\mathit{getEventHandlers}}
\newcommand{\ag}{\mathit{arg}}
\newcommand{\yield}{\mathit{isPreemptionPoint}}
\newcommand{\Yield}{\mbox{preemption}}
\newcommand{\R}{\mathcal{R}}

\newcommand {\veq}{\approx^{\beta}_L}
\newcommand {\sigeq}{\simeq^{\beta}_L}
\newcommand {\teq}{\approx^{\beta}_L}
\newcommand {\req}{\cong^{\beta}_L}
\newcommand {\sceq}{\sim^{\beta}_L}
\newcommand {\lsd}{\Gamma(!\sigma(\dst))}
\newcommand{\eq}{\sim}

\newcommand{\Figure}[1]{Fig.~\ref{#1}}
\newcommand{\TT}[1]{\lstinline[basicstyle=\ttfamily\normalsize]{#1}}%\def\TT\lstinline

\renewcommand{\lstlistingname}{Listing}
\renewcommand{\thelstlisting}{\arabic{lstlisting}}
\lstset{
  numbers=left,
  numbersep=10pt,
  xleftmargin=2em,
  basicstyle=\ttfamily\small,
  escapeinside={@}{@},
  keywordstyle=\ttfamily,
  firstnumber=1,
  showspaces=false,
  belowskip=-0.5em,
  numberfirstline=true,
  columns=fullflexible,
  captionpos=b,
  breaklines=true,
  mathescape=true,
  tabsize=2
}

\hyphenation{non-inter-fer-ence}
\hyphenation{Java-Script}

\title{WebPol: Fine-grained Information Flow Policies for Web
  Browsers}
\titlerunning{WebPol}

\author{Abhishek Bichhawat\inst{1} \and Vineet Rajani\inst{2} \and
  Jinank Jain\inst{3} \and Deepak Garg\inst{2} \and Christian
  Hammer\inst{4}}
\institute{Saarland University, Germany
\and MPI-SWS, Germany
\and ETH Zürich, Switzerland
\and University of Potsdam, Germany}

\maketitle

% Use the following at camera-ready time to suppress page numbers.
% Comment it out when you first submit the paper for review.
\thispagestyle{empty}
\ifconf{
\pagenumbering{gobble}
}
\setcounter{footnote}{0}

%!TEX root = ifc-policies16.tex

\begin{abstract}
  In the standard web browser programming model, third-party scripts
  included in an application execute with the same privilege as the
  application's own code. This leaves the application's confidential
  data vulnerable to theft and leakage by malicious code and
  inadvertent bugs in the third-party scripts. Security mechanisms in
  modern browsers (the same-origin policy, cross-origin resource
  sharing and content security policies) are too coarse to suit this
  programming model. All these mechanisms (and their extensions)
  describe whether or not a script can access certain data, whereas
  the meaningful requirement is to allow untrusted scripts access to
  confidential data that they need and to prevent the scripts from
  leaking data on the side.  Motivated by this gap, we propose \sys, a
  policy mechanism that allows a website developer to include
  fine-grained policies on confidential application data in the
  familiar syntax of the JavaScript programming language. The policies
  can be associated with any webpage element, and specify \emph{what}
  aspects of the element can be accessed \emph{by which third-party
    domains}. A script can access data that the policy allows it to,
  but it cannot pass the data (or data derived from it) to other
  scripts or remote hosts in contravention of the policy. To specify
  the policies, we expose a small set of new native APIs in
  JavaScript.
%
%  We describe the policy
%  model and provide several illustrative examples of its use in
%  practice.
%
  Our policies can be enforced using any of the numerous existing
  proposals for information flow tracking in web browsers. We have
  integrated our policies into one such proposal that we use to
  evaluate performance overheads and to test our examples.

\end{abstract}

\section{Introduction}
\label{sec:intro}

Webpages today rely on third-party JavaScript to provide useful
libraries, page analytics, advertisements and many other
features. JavaScript works on a \emph{mashup} model, wherein the
hosting page and included scripts share the page's state (called the
DOM). Consequently, by design, all included third-party scripts run
with the same access privileges as the hosting page. While some
third-party scripts are developed by large, well-known, trustworthy
vendors, many other scripts are developed by small, domain-specific
vendors whose commercial motives do not always align with those of the
webpage providers and users. This leaves sensitive information such as
passwords, credit card numbers, email addresses, click histories,
cookies and location information vulnerable to inadvertent bugs and
deliberate exfiltration by third-party scripts. In many cases,
developers are fully aware that a third-party script accesses
sensitive data to provide useful functionality, but they are unaware
that the script also leaks that data on the side. In fact, this is a
widespread problem~\cite{jang10CCS}.

%% Additionally, these scripts have
%% uncontrolled access to other private information of the users related
%% to the host page that includes these scripts like cookies, location
%% etc.

Existing web security standards and web browsers address this problem
unsatisfactorily, favoring functionality over privacy. The same-origin
policy (SOP)~\cite{sop} implemented in all major browsers restricts a
webpage and third-party scripts included in it to communicating with
web servers from the including webpage's domain only. However, broad
exceptions are allowed. For instance, there is no restriction on
request parameters in urls that fetch images and, unsurprisingly,
third-party scripts leak information by encoding it in image urls. The
candidate web standard Content Security Policy (CSP)~\cite{csp}, also
implemented in most browsers, allows a page to white list scripts that
may be included, but places no restriction on scripts that have been
included, thus not helping with the problem above. Other mechanisms
(including a provision in the SOP) restrict scripts loaded in a
different third-party window or frame from accessing the resources of
a page but do not restrict third-party scripts included in the page
itself.

The academic community has recently proposed solutions based on
information flow control
(IFC)~\cite{flowfox,csf14,jsflow,post14,csf15,chudnov-ccs,cowl,jang10CCS},
also known as mandatory access control. Their \emph{ideal} goal is to
allow third-party scripts access to necessary sensitive data, but
restrict \emph{where} scripts can send the data---and data derived
from that data---in accordance with a policy. While this would balance
functionality and privacy perfectly, all existing IFC-based solutions
for web browsers fall short of this ideal goal. Many proposals,
including several taint-based
solutions~\cite{jsflow,post14,csf15,chudnov-ccs}, focus on the IFC
mechanism, but currently lack adequate support for specifying policies
conveniently. Flowfox~\cite{csf14} provides a rich policy framework
but all websites are subject to the same policy, and the underlying
IFC technique, secure multi-execution~\cite{SME}, does not handle
shared state soundly. COWL~\cite{cowl} uses coarse-grained isolation,
allowing scripts' access to either remote domains or the shared state,
but not both. This requires significant code changes when both are
needed simultaneously (see Section~\ref{sec:related} for more
details).

The contribution of our work is {\sys}, a policy framework that allows
a webpage developer to release data selectively to third-party scripts
(to obtain useful functionality), yet control what the scripts can do
with the data. {\sys} integrates with any taint-based IFC solution to
overcome the shortcomings listed above. {\sys} policies label
sensitive content (page elements and user-generated events) at source,
and selectively declassify them by specifying where (to which domains)
the content and its derivatives can flow. Host page developers specify
{\sys} policies in JavaScript, a language already familiar to them.

% Labels are context-sensitive---they can depend on the current state of the
% application. {\sys} policies are specified in privileged JavaScript
% event handlers by the developers of the hosting page. Both JavaScript
% and event handlers are already familiar to web developers.

Under the hood, any taint-based IFC solution can be used to track data
flows and to enforce {\sys} policies. As a demonstrative prototype, we
have integrated {\sys} with our previous taint-based IFC framework for
WebKit~\cite{post14,csf15}, the engine that powers Apple's Safari and
other browsers. We demonstrate the expressiveness of {\sys} policies
through examples and by applying {\sys} to two real websites. Through
measurements, we demonstrate that {\sys} policies impose
low-to-moderate overhead, which makes {\sys} usable today.
\iffull{Through a
small lab study (described in
  Appendix~\ref{sec:user-study}), we test that {\sys} can be
effectively used by programmers familiar with HTML and JavaScript.}
\ifconf{A full version of the paper~\cite{fullversion-esorics17}
  contains more details, including a small lab study through which we
  test that {\sys} can be effectively used by programmers familiar
  with HTML and JavaScript.}

\section{Overview}
\label{sec:overview}

This section provides an overview of information flow control (IFC) in
the context of web browsers and lists important considerations in the
design of {\sys}. IFC is a broad term for techniques that control the
flow of sensitive information in accordance with pre-defined
policies. Sensitive information is information derived from sources
that are confidential or private. Any IFC system has two
components---the \emph{policy component} and the \emph{enforcement
  component}. The policy component allows labeling of private
information sources. The label on a source specifies how private
information from that source can be used and where it can flow. The
collection of rules for labeling is called the policy. The enforcement
component enforces policies. {\sys} contributes a policy component to
complement existing work on enforcement components in web
browsers. Many existing enforcement components can be used with
{\sys}. For completeness, we describe both policy and enforcement
components here.

\medskip 
\noindent \textbf{Policy component.}  The policy component provides a
way to label or mark sensitive data sources with labels that represent
confidentiality and where data can flow. In the context of webpages,
data sources are objects generated in response to user events like the
content of a password box generated due to key presses or a mouse
click on a sensitive button, and data obtained in a network receive
event. In {\sys}, data sources can be labeled with three kinds of
labels, in increasing order of confidentiality: 1) the label
\texttt{public} represents non-sensitive data, 2) for each domain
\texttt{domain}, the label \texttt{domain} represents data private to
the domain; such data's flow should be limited only to the browser and
servers belonging to \texttt{domain} and its subdomains, and 3) the
label \texttt{local} represents very confidential data that must never
leave the browser. Technically, labels are elements of the partial
order \texttt{public} $<$ \texttt{domain}$_i$ $<$
\texttt{local}. Labels higher in the order represent more
confidentiality than labels lower in the order. These labels are
fairly expressive.\footnote{Richer label models that support, for
  instance, conjunctions and disjunctions of labels~\cite{dc-labels}
  are compatible with {\sys}. However, we have not found the need for
  such models so far.} For example, labeling a data source with the
domain of the hosting page prevents exfiltration to
third-parties. Labeling a data source with the domain of a third-party
provider such as an page analytics provider allows transfer to only
that service.

Since most data on a webpage is not sensitive, it is reasonable to
label data sources \texttt{public} by default and only selectively
assign a different label. {\sys} uses this blacklisting approach.  Two
nuances of source labeling are noteworthy. The first is its fine
granularity. Not all objects generated by the same class of events
have the same label. For instance, characters entered in a password
field may have the domain label of the hosting page, limiting their
flow only to the host, but characters entered in other fields may be
accessible to third-party advertising or analytics scripts without
restrictions. This leads to the following requirement on the policy
component.

\medskip
\noindent \textit{Requirement 1:} The policy component must allow
associating different policies with different elements of the page.
\medskip

The second nuance is that the label of an object can be dynamic, i.e.,
history-dependent. Consider a policy that hides from an analytics
script how many times a user clicked within an interactive panel, but
wants to share whether or not the user clicked at least once. The
label of a click event on the panel is \texttt{public} the first time
the user clicks on it and private afterwards and, hence, it depends on
the history of user interaction.
% (it suffices to track a boolean value in this case)
This yields the following requirement on the policy
component.

\medskip
\noindent \textit{Requirement 2:} Labels may be determined
dynamically. This requirement means that labels must be set by
\emph{trusted policy code} that is executed on-the-fly and that has
local state.
\medskip

%% Section~\ref{sec:model} explains how {\sys} satisfies these
%% requirements.

%% {\sys} overloads the web browser's existing event handler mechanism
%% to allow the specification of policies satisfying Requirements~1
%% and~2.
% We explain this in Section~\ref{sec:model}.

\medskip \noindent \textbf{Enforcement component.}  Source data labels
must be enforced even as scripts transform and transmit the
data. Existing literature is rife with techniques for doing this, even
in the context of web-browsers. Fine-grained taint
tracking~\cite{jang10CCS,jsflow,post14,csf15,chudnov-ccs},
coarse-grained taint tracking~\cite{cowl,ndss15}, secure
multi-execution~\cite{SME,csf14}, faceted
execution~\cite{austin12POPL,jeeves} and static
analysis~\cite{guarnieri11ISSTA,stagedIFC,mashif} are some enforcement
techniques that have been considered in the context of
JavaScript. They differ considerably in their mechanics, their
expressiveness and ease of fit with the browser programming
model. {\sys} has been designed keeping fine-grained taint tracking
(FGTT) in mind, so we explain that technique in some detail below.

In FGTT, the language runtime is modified to track information flows
and to attach a label (often called a taint) with each runtime object,
including objects on the stack, the heap and, in the context of web
browsers, the DOM.  Two kinds of flows are typically
considered. \emph{Explicit} flows arise as a result of direct
assignment. In these cases, the label of the destination object is
overwritten with the label of the source object. \emph{Implicit} flows
arise due to control dependencies. For instance, in \texttt{pub =
  false; if (sec) pub = true}, the final value of \texttt{pub} depends
on \texttt{sec} although there is no direct assignment from
\texttt{sec} to \texttt{pub}. Implicit flows are tracked by keeping a
\emph{context} label on the instruction pointer. Once both explicit
and implicit flows are tracked, enforcing policies is straightforward:
An outgoing communication with domain \texttt{d}'s servers is allowed
only if the labels on the payload of the communication and the
instruction pointer at the point of the communication are either
\texttt{public} or~\texttt{d}. This ensures that all labels attached
to source data are respected.

FGTT can be implemented either by modifying the browser's JavaScript
engine to track flows and labels~\cite{jsflow,post14}, or by a
source-to-source transform of JavaScript code prior to
execution~\cite{chudnov-ccs}. There is a space and time overhead
associated with storing labels and tracking them. However, with
careful engineering, this overhead can be reduced enough to not be
noticeable to end-users.

\lstset{language=HTML}

\section{{\sys} policy model}
\label{sec:model}
{\sys} works on a browser that has already been augmented with IFC
enforcement.
% Such an enforcement labels all objects internally.
It provides a framework that allows setting labels at
fine-granularity, thus expressing and enforcing rich policies. This
section describes the threat model for {\sys} and explains the {\sys}
design.

\medskip \noindent \textbf{Threat model.}
{\sys} prevents under-the-hood exfiltration of sensitive data that has
been provided to third-party scripts for legitimate reasons. So,
third-party scripts are not trusted but code from the host domain is
trusted.

%% In fact, {\sys} policies are included in the host page's HTML
%% (explained below) and we assume that the policies are written
%% correctly.

We are interested only in JavaScript-level bugs or exfiltration
attempts. We trust the browser infrastructure to execute all
JavaScript code following the language's semantics and to dispatch
events correctly. Low-level attacks that target vulnerabilities in the
browser engine are out of scope. Similarly, defending against network
attacks (like man-in-the-middle attacks) is not our goal. Orthogonal
techniques like end-to-end encryption or HTTPS can be used to defend
against those attacks. Integrity attacks are also out of scope. For
instance, attacks based on sending requests containing no sensitive
data to websites, where the user might already be logged in, cannot be
prevented using this model.

{\sys} executes on top of an IFC enforcement in the browser. That
enforcement is assumed to be correct and to track all flows. Prior
work on such enforcement has often been supplemented with formal
proofs to show that the enforcement is correct, at least
abstractly~\cite{jsflow,post14,csf15}.

{\sys}'s policies are agnostic to specific channels of information
leak.  However, current IFC enforcements in browsers track only
explicit and implicit flows. Consequently, leaks over other channels
such as timing and memory-usage are currently out of scope. As IFC
enforcements improve to cover more channels, {\sys}'s policies will
extend to them as well.

\subsection{Policies as event handlers}

The first question in the design of {\sys} is who should specify
policies. Since our goal is to prevent exfiltration of data by
third-party scripts and it is the developer of the host page who
bootstraps the inclusion of scripts and best understands how data on
the page should be used, it is natural and pragmatic to have the
developer specify policies, possibly as part of the page itself.

The next question is how the developer specifies policies. To answer
this, we recall the two requirements we identified in
Section~\ref{sec:overview}---it should be possible to specify
different policies on different page elements and policies should be
allowed to include code that is executed on-the-fly to generate
labels. When we also consider the fact that sensitive data is usually
generated by input events, it is clear that policies should \emph{be}
page element-specific, (trusted) code that is executed after events
have occurred (this code labels event-generated data). Fortunately,
web browsers provide exactly this abstraction in the form of event
handlers! So, we simply extend the event-handling logic in web
browsers to express {\sys} policies. This allows us to leverage a lot
of the existing browser logic for event handler installation, parsing
and event dispatch. Before explaining how we do this, we provide a
brief overview of event handling in web browsers.

\medskip \noindent \textbf{Event handlers and event dispatch.}
Browsers execute JavaScript functions, called event handlers, in
response to input events like mouse clicks, key presses, and
asynchronous network receives. Save for network receive events, every
event has a \emph{target}, which is an element in the page's DOM where
the event originated. For instance, if a button is clicked, the target
of the ensuing event is the button. Code running on a page can add an
event handler on any element on the page, listening for a specific
event. When an event occurs, all handlers associated for that event
with the event's target and the target's ancestors are triggered
sequentially. This is called \emph{event dispatch.} The specific order
in which handlers are triggered is not relevant for our purposes
(although it is fairly interesting for IFC
enforcement~\cite{csf15}). The whole process is bootstrapped by the
static HTML of the page, which may contain JavaScript that is executed
when the page loads initially, and this JavaScript installs the first
set of event handlers.

%% There are two salient aspects of event dispatch that are relevant for
%% {\sys}. First, there is no specific order in which event handlers
%% registered for an event on an element are triggered. Second, when an
%% event occurs on an element, event handlers for that event associated
%% with all \emph{ancestors} of that element (i.e., that elements
%% enclosing elements in the page rendering) are also triggered. The
%% logic for doing this is complex and largely irrelevant for our
%% purposes, but the important aspect is that

\medskip \noindent \textbf{Policy handlers.}
In {\sys}, policies are special event handlers, specified using a
special marker in the HTML source of the hosting page. These special
handlers, called \emph{policy handlers}, follow standard JavaScript
syntax, can be attached to any page element, listening for any event
and, like other handlers, are triggered every time the event is
dispatched on the element or any of its descendants in the
DOM. However, unlike other handlers, the sole goal of policy handlers
is to assign labels to other sensitive objects, including the event
being dispatched. To allow the policy handlers to do this, we modify
the browser slightly to afford these handlers two special privileges:
\begin{itemize}
\item Policy handlers can execute two new JavaScript API functions
  that set labels on other objects. No other JavaScript code can
  execute these two functions. These functions are described later.
\item During event dispatch all applicable policy handlers are executed
  before ordinary handlers. This ensures that labels are set before
  ordinary handlers (including those of third-party scripts) execute.
\end{itemize}
To maintain the integrity of the policies, policy handlers must be
included in the HTML source of the page directly. They \emph{cannot}
be installed dynamically by JavaScript code. Otherwise, third-party
scripts could install policy handlers that set very permissive labels.
Also, if a DOM element has a policy handler, we disallow third-party
scripts from detaching that element or moving it elsewhere, as that
can change the interpretation of the policy. Similarly, changing the
attributes of such an element is restricted.

%% (A third-party script can, however, use transparent overlays to
%% steal infromation. We show how to specify a policy that protects
%% against such attacks in the Appendix.)

\begin{figure}[tb]
  \centering \includegraphics[width=7cm]{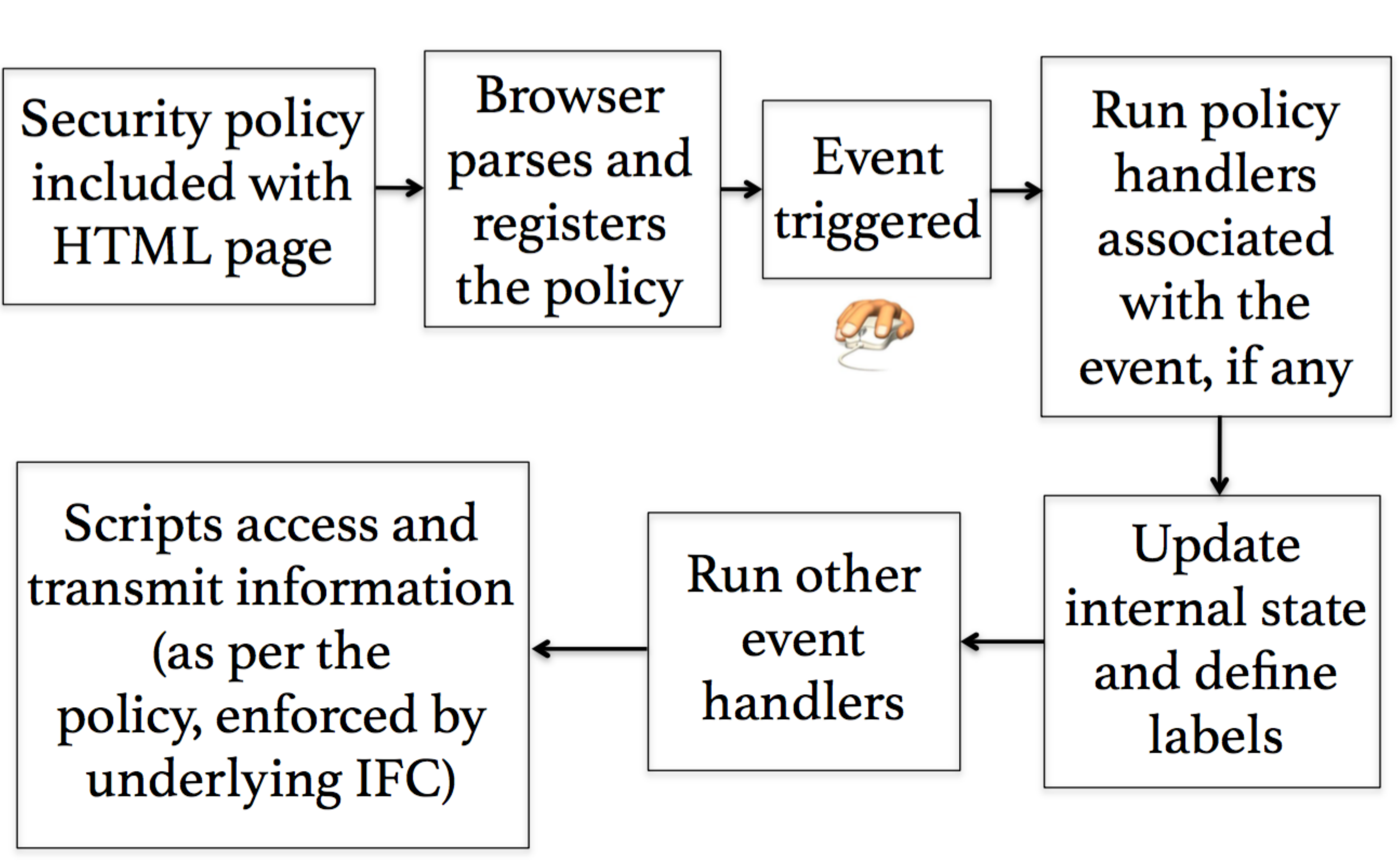} \caption{Workflow
  of the {\sys} policy model} \label{fig:model}
\end{figure}

Since different policy handlers can be associated with different
elements, Requirement~1 is satisfied.  Moreover, policy handlers are
ordinary JavaScript code, so they can also maintain local state in
private variables, thus satisfying Requirement~2. 

The workflow of policy interpretation in {\sys} is shown in
Figure~\ref{fig:model}. We briefly summarize the steps:

\begin{enumerate}
\item The web page developer specifies the policy in the host HTML
  page in the form of special event handlers.
\item The browser parses the policy and registers its handlers (mostly
  like usual handlers, but with the two special privileges mentioned
  above).
\item When an event dispatches, listening policy handlers are
  executed first.
\item These policy handlers set labels on objects affected by the
  event, including the event object itself. They may also
  update any local state they maintain.
\item The remaining event handlers are dispatched as usual.
  The IFC enforcement in the browser enforces all labels that have
  been set by the policy handlers (during any prior event's dispatch),
  thus preventing any data leak in contravention of the labels.
\end{enumerate}

\subsection{Integration with the web browser}

% \begin{lstlisting}[float, caption=Policy inclusion on a web page,label=egps]
% <!-- hostpage.html -->
% ...
% <script src="policyfile.policy"> </script> 
% ...
% \end{lstlisting}

{\sys} needs minor modifications to the browser to parse and interpret
policies and to expose additional JavaScript API functions to set
labels.

\medskip \noindent \textbf{HTML and event dispatch changes.}
{\sys} adds an HTML extension to differentiate policy code from other
JavaScript code. Concretely, we change the browser's parser to
interpret any script file with the extension \texttt{.policy} included
directly in the host page as a policy. If such a \emph{policy script}
installs a handler, it is treated as a policy handler. Additionally, a
policy script can set labels on the page's global variables and DOM
elements (like password fields). If a script does this, it should be
included in the host page before third-party scripts that use those
variables. %% A policy script is included as:
%% $$\texttt{<script src="policyfile.policy"></script>}$$
%
{\sys} also requires a small change to the browser's event dispatch
mechanism to execute policy handlers before other handlers.

%% It is important to include policy scripts before third-party
%% scripts in the host page.

%% The policy starts exceutes only when the policy script is parsed by
%% the browser. Thus, the placement of the policy script on the web page
%% is important to ensure that the policy executes before any
%% unintentional data leak occurs.

\medskip \noindent \textbf{Label-setting APIs.}
{\sys} exposes two new JavaScript API functions to set labels. These
functions can be called only by the policy code in \texttt{.policy}
files and handlers installed by such files (we modify the browser to
enforce this).

The function \texttt{setLabel(label)} sets the label of the object on
which it is called to \texttt{label}. As explained
earlier, \texttt{label} can be \texttt{public}, a domain name,
or \texttt{local} (the default is \texttt{public}). Once an object's
label is set, it is enforced by the underlying IFC enforcement. The
special label \texttt{HOST} is a proxy for the domain of the host
page.

The function \texttt{setContext(label)} can be called only on an event
object. It restricts the visibility of the event to
label \texttt{label} and higher. In simple terms, if \texttt{label} is
a domain, then only that domain can ever learn that this event
occurred, whereas if \texttt{label} is \texttt{local}, then no domain
can ever learn that this event occurred. Technically, this is
accomplished by setting the so-called $\PC$ or program counter label
of event handlers running during the dispatch to \texttt{label}, which
ensures that their side-effects (writes to DOM and network
communication) are labeled \texttt{label} or higher.

As opposed to \texttt{setLabel}, which makes individual data objects
(like password fields) private, \texttt{setContext} makes the
\emph{existence} of an event private. This is useful.
For instance, clicking on the ``politics'' section of a news feed
might indicate that the user is interested in politics, which may be
private information, so the page may want to hide even the existence
of click events from third-party scripts. (The distinction between the
privacy of event content and event occurrence has been previously
described by Rafnsson and Sabelfeld~\cite{Rafnsson-csf13}.)

\section{Examples}
\label{sec:examples}

We illustrate the expressiveness of {\sys} policies through a few
examples.

\begin{lstlisting}[float, caption=Password strength checking script that leaks the password,label=egscript1,language=C,escapechar=\%]
var p = document.getElementById("pwd");
p.addEventListener("keypress", function (e){%\label{pwd:listener}%
  var score = checkPwdStrength(p.value); %\label{pwd:check}%
  document.getElementById("pwdStrength").innerText = score; %\label{pwd:score}%
  new Image().src = "http://stealer.com/pwd.jsp?pwd="+p +score; %\label{pwd:req}%
});
\end{lstlisting}

% \begin{lstlisting}[float, caption=Policy for protecting password,label=egpol1]
% document.getElementById("pwd").setLabel("login.com");  
% \end{lstlisting}

\medskip
\noindent \textbf{Example 1: Password strength checker.}  Many
websites deploy \emph{password strength checkers} on pages where users
set new passwords. A password strength checker is an event handler
from a third-party library that is triggered each time the user enters
a character in the new password field. The handler provides visual
feedback to the user about the strength of the password entered so
far. Strength checkers usually check the length of the password and
the diversity of characters used. Consequently, they do not require
any network communication. However, standard browser policies cannot
enforce this and the password strength checker can easily leak the
password if it wants to. Listing~\ref{egscript1} shows such a
``leaky'' password checker. The checker installs a listener for
keypresses in the password field (line~\ref{pwd:listener}). In
response to every keypress, the listener delivers its expected
functionality by checking the strength of the password and indicating
this to the user (lines~\ref{pwd:check},~\ref{pwd:score}), but then it
leaks out the password to \texttt{stealer.com} by 
requesting an image at a url that includes the password
(line~\ref{pwd:req}). The browser's standard SOP allows this.

With {\sys}, the developer of the host webpage can prevent any
exfiltration of the password by including the policy script:

\medskip
\texttt{document.getElementById("pwd").setLabel("HOST");}

\medskip 
% shown in Example~\ref{egpol1}. 
\noindent This policy sets the label of the password field to the
host's own domain using the function
\texttt{setLabel()}. Subsequently, the IFC enforcement restricts all
outgoing communication that depends on the password field to the host.

Conceptually, this example is simple because it does not really
leverage the fine-granularity of {\sys} policies and FGTT. Here, the
third-party script does not need any network communication for its
intended functionality and, hence, simpler confinement mechanisms that
prohibit a third-party script from communicating with remote servers
would also suffice. Our next example is a scenario where the
third-party script legitimately needs remote communication. It
leverages the fine-granularity of {\sys} policies and FGTT.

\begin{lstlisting}[float, caption=Currency converter script that leaks a private amount,label=egcc,escapechar=\%]
function currencyConverter() {
	var toCur = document.getElementById("to").value;
	var xh = new XMLHttpRequest();
	xh.onreadystatechange = function() { %\label{startcallback}%
		if (xh.readyState == 4) { %\iffalse && xhttp.status == 200 \fi %
			currencyRate = eval(xhttp.responseText);%\label{startsend}%
			var aAmt = document.getElementById("amt").value;
			var convAmt = aAmt * currencyRate;
			document.getElementById("camt").innerHTML = convAmt;%\label{ressend}%
			xh.open("GET","http://currConv.com/amount.jsp?atc=" + aAmt);%\label{endsend}%
			xh.send(); }} %\label{endcallback}%
	xh.open("GET","http://currConv.com/conv.jsp?toCur=" + toCur, true);
	xh.send(); }%\label{mainsend}%
\end{lstlisting}

\medskip
\noindent \textbf{Example 2: Currency conversion.}
Consider a webpage from an e-commerce website which displays the cost
of an item that the user intends to buy. The amount is listed in the
site's native currency, say US dollars (USD), but for the user's
convenience, the site also allows the user to see the amount converted
to a currency of his/her choice. For this, the user selects a currency
from a drop-down list. A third-party JavaScript library reads both the
USD amount and the second currency, converts the amount to the second
currency and inserts it into the webpage, next to the USD amount.
The third-party script fetches the current conversion rate from its
backend service at \texttt{currConv.com}. Consequently, it must send
the \emph{name} of the second currency to its backend service, but
must not send the amount being converted (the amount is private
information). The web browser's same-origin policy has been relaxed
(using, say, CORS~\cite{cors}) to allow the script to talk to its
backend service at \texttt{currConv.com}. The risk is that the script
can now exfiltrate the private amount. Listing~\ref{egcc} shows a
leaky script that does this. On line~\ref{mainsend}, the script makes
a request to its backend service passing to the second currency. The
callback handler (lines \ref{startcallback}--\ref{endcallback}) reads
the amount from the page element \texttt{amt}, converts it and inserts
the result into the page (lines
\ref{startsend}--\ref{ressend}). Later, it leaks out the amount to the
backend service on line~\ref{endsend}, in contravention of the
intended policy.

With {\sys}, this leak can be prevented with the following policy that
sets the label of the amount to the host only:

\medskip
\texttt{document.getElementById("amt").setLabel("HOST")}

\medskip
\noindent This policy prevents exfiltration of the amount but does not
interfere with the requirement to exfiltrate the second
currency. Importantly, no modifications are required to a script that
does not try to leak data (e.g., the script obtained by dropping the
leaky line~\ref{endsend} of Listing~\ref{egcc}).

\begin{lstlisting}[float,caption=Policy that allows counting clicks but hides details of the clicks,label=eganal1,escapechar=\%]
var p = document.getElementbyId("sect_name");
p.addEventListener("click",function(event){
  event.setLabel("HOST"); });
\end{lstlisting}

\medskip
\noindent\textbf{Example 3: Web analytics.} To better understand how
users interact with their websites, web developers often include
third-party analytics scripts that track user clicks and keypresses to
generate useful artifacts like page heat-maps (which part of the page
did the user interact with the most?). Although a web developer might
be interested in tracking only certain aspects of their users'
interaction, the inclusion of the third-party scripts comes with the
risk that the scripts will also record and exfiltrate other private
user behavior (possibly for monetizing it later). Using {\sys}, the
web developer can write precise policies about which user events an
analytics script can access and when. We show several examples of
this.

\begin{lstlisting}[float, caption=Analytics script that counts
  clicks,label=egan,escapechar=\%]
clickCount = 0;
var p = document.getElementbyId("sect_name");  
p.addEventListener("click",function clkHdlr(e){ clickCount += 1; });
\end{lstlisting}

To allow a script to only count the number of occurrences of a class
of events (e.g., mouse clicks) on a section of the page, but to hide
the details of the individual events (e.g., the coordinates of every
individual click), the web developer can add a policy handler on the
top-most element of the section to set the label of the individual
event objects to \texttt{HOST}. This prevents the analytics script's
listening handler from examining the details of individual events, but
since the handler is still invoked at each event, it can count their
total number. Listings~\ref{eganal1} and~\ref{egan} show the policy
handler and the corresponding analytics script that counts clicks in a
page section named \texttt{sect\_name}.

%% \begin{lstlisting}[float,caption=Policy allowing selected access,label=egpol3]
%% window.addEventListener("click",function(event){
%%     event.setLabel("host.com");
%%     event.setContext("analytics.com");
%% });
%% \end{lstlisting}

%% \begin{lstlisting}[float, caption=Modified analytics script as per policy,label=eganp]
%% window.addEventListener("click",function clkHdlr(e){
%%     var coord = new Image(); 
%%     coord.src ="http://analytics.com/tracker.jsp?click=true";
%% });
%% \end{lstlisting}

\begin{lstlisting}[float,caption=Policy that tracks whether a click
  happened or not only,label=eganal2,escapechar=\%]
var alreadyClicked = false;
var p = document.getElementById("sect_name");
p.addEventListener("click",function(event){
  if (alreadyClicked = true) event.setContext("HOST"); %\label{click:cont}%
  else {alreadyClicked = true; event.setLabel("HOST");} }); %\label{click:lab}%
\end{lstlisting}

Next, consider a restriction of this policy, which allows the
analytics script to learn only whether or not \emph{at least one}
click happened in the page section, completely hiding clicks beyond
the first. This policy can be represented in {\sys} using a local
state variable in the policy to track whether or not a click has
happened. Listing~\ref{eganal2} shows the policy. The policy uses a
variable \texttt{alreadyClicked} to track whether or not the user has
clicked in the section. Upon the user's first click, the policy
handler sets the event's label to the host's domain
(line~\ref{click:lab}). This makes the event object private but
allows the analytics handler to trigger and record the occurrence of
the event. On every subsequent click, the policy handler sets the
event's \emph{context} to the host domain using \texttt{setContext()}
(line~\ref{click:cont}). This prevents the analytics script from
exfiltrating any information about the event, including the fact that
it occurred. 

Finally, note that a developer can subject different page sections to
different policies by attaching different policy handlers to them. The
most sensitive sections may have a policy that unconditionally sets
the event context to the host's, effectively hiding all user events in
those sections. Less sensitive sections may have policies like those
of Listings~\ref{eganal2} and~\ref{eganal1}. Non-sensitive sections
may have no policies at all, allowing analytics scripts to see all
events in them.

\medskip
\noindent \textbf{Example 4: Defending against overlay-based attacks.}
\iffull{Appendix~\ref{sec:overlay} describes}\ifconf{The full version
  of the paper~\cite{fullversion-esorics17} describes} a simple {\sys}
policy that defends against an attack where a malicious script creates
a transparent overlay over a sensitive element (like a password field)
to record user events like keypresses without policy protection.

\iffull{
\medskip
\noindent \textbf{Summary of {\sys} expressiveness.}
We end this
section by commenting on the expressiveness of {\sys} policies in
broad terms. The security community has extensively studied several
aspects of policy labeling, colloquially called the \emph{dimensions
  of declassification} (see~\cite{dimDecl} for a survey). Broadly
speaking, {\sys} policies cover three of these dimensions---the
policies specify what data is declassified (dimension: what), to which
domains (dimension: to whom) and under what state (dimension:
when). ``What data is declassified'' is specified by selectively
attaching policies to elements of the page. ``Which domains get
access'' is determined directly by the labels that the policy
sets. Finally, labels generated by policy handlers can depend on
state, as illustrated in Listing~\ref{eganal2}.

There are two other common dimensions of labeling---who can label the
data (dimension: who) and where in the code can the labels change
(dimension: where). These dimensions are fixed in {\sys} due to the
specifics of our problem: All policies are specified by the host page
in statically defined policies.
}

%% \subsection{Declassification dimensions}

%% \paragraph{What specification}
%% The policy can specify \emph{what} data (generally, input) is secret
%% or private, and \emph{what part} of the secret data can be
%% released. The policy code has access to special APIs that can tag or
%% label a JavaScript object or an HTML element based on the
%% specification by the developer. The policy can also specify if the
%% occurrence of an  event like mouse-click or keypress on a particular
%% HTML element is secret or is selectively  visible to some third-party
%% scripts.

%% \paragraph{Whom specification}
%% The policy can specify \emph{which} third-party scripts or domains
%% have access to the released information by labeling it appropriately,
%% i.e., with the label of the domain to which the access is allowed.

%% \paragraph{When specification}
%% The policy can specify \emph{when} some secret information can be
%% released (generally, based on the occurrence of some event). This is
%% specified as a JavaScript event handler, and thus, can be associated
%% with any DOM node or HTML element (e.g., window, div, input)
%% on the page for any event (e.g., keypress, click). If an HTML element
%% \emph{n} does not have a policy registered for an event \emph{e}, the
%% policy registered with the closest parent (ancestor) of the element
%% \emph{n} for the event \emph{e} is enforced. The host page can also
%% specify some intialization policies that execute before any input
%% (event) is received from the user. Only one policy is enforced per
%% event; thus, the policy closest to the element is enforced.

\section{Implementation}
\label{sec:impl}

We have prototyped {\sys} in WebKit, a popular open source browser
engine that powers many browsers including Apple's Safari. Our
implementation runs on top of our prior IFC enforcement in WebKit that
uses FGTT and a bit of on-the-fly static analysis~\cite{post14}. The
IFC enforcement is highly optimized, and covers most JavaScript native
functions (the DOM API)~\cite{csf15}. It targets WebKit nightly build
$\#r122160$ and works with the Safari web browser, version 6.0.
% \textcolor{red}
{Since it is difficult to port our earlier
  implementation (not {\sys}) to a newer version of WebKit, we choose
  to evaluate {\sys} on this slightly outdated setup. This suffices
  since {\sys}'s design is not affected by recent browser updates.}
The source code is available online at:
\url{https://github.com/bichhawat/ifc4bc}. 

Our earlier IFC implementation modified approximately 6,800 lines in
the JavaScript engine, the DOM APIs and the event handling logic for
FGTT.
To implement {\sys}, we additionally modified the HTML parser to
distinguish policy files (extension \texttt{.policy}) from other
JavaScript files and to give policy code extra privileges. We also
added the two new JavaScript API functions \texttt{setLabel()} and
\texttt{setContext()}. Finally, we modified the event dispatch logic
to trigger policy handlers before other handlers. In all, we changed
25 lines in the code of the parser, added 60 lines for the two new API
functions and changed 110 lines in the event dispatch logic. Overall,
implementing {\sys} has low overhead, and we expect that it can also
be ported to other browsers or later versions of WebKit easily.

\section{Evaluation}
\label{sec:eval}

The goal of our evaluation is two-fold. First, we want to measure
{\sys} overhead, both on parsing and installing policies during page
load and on executing policy handlers later. We do this for four
examples presented in Section~\ref{sec:examples} and for two
real-world websites. Second, we wish to understand whether {\sys} can
be used easily. Accordingly, we apply {\sys} policies to two
real-world websites and report on our experience.
\iffull{We also conducted a small user-study where we asked
  programmers not already familiar with {\sys} to write {\sys}
  policies for the examples of Section~\ref{sec:examples} after a
  brief tutorial introduction. We relegate our observations from the
  user-study to Appendix~\ref{sec:user-study} and focus here only on
  performance overheads and the application to real-world websites.}
All our experiments were
performed on a 3.2GHz Quad-core Intel Xeon processor with 8GB RAM,
running Mac OS X version 10.7.4 using the browser configuration
described in Section~\ref{sec:impl}.
%
% \TODO{Fix this machine configuration. I think
%  this is not accurate any more.}

\medskip\noindent\textbf{Performance overheads on synthetic examples.}
To measure {\sys}'s runtime overhead, we tested four examples from
Section~\ref{sec:examples} (Examples~1,~2 and the two sub-examples of
Example~3) in three different configurations:
\textbf{Base}---uninstrumented browser, no enforcement;
\textbf{IFC}---taint tracking from prior work, but no policy handlers
(everything is labeled public); \textbf{{\sys}}---our system running
policy handlers and taint tracking.

\newcolumntype{C}[1]{>{\centering\let\newline\\\arraybackslash\hspace{0pt}}m{#1}}

\begin{table}[tbp]
\centering
\begin{tabular}{ | C{2cm} || C{1.5cm} | C{1.5cm} | C{1.5cm} || C{1.5cm} | C{1.5cm} | C{1.5cm} |}
\hline
\multicolumn{1}{|c||}{ } &
\multicolumn{3}{ c ||}{JavaScript Execution Time} &
\multicolumn{3}{ c |}{Page Load Time} \\
\hline
 Example \# & \textbf{Base} & \textbf{IFC} & \textbf{\sys} & \textbf{Base} & \textbf{IFC} & \textbf{\sys} \\
  \hhline{|=#=|=|=#=|=|=|} 
  Example~1 & 2430 & 2918 (+20.1\%) & 2989 (+1.9\%) & 16 & 17 (+6.3\%) & 19 (+12.5\%) \\  
\hline
  Example~2 & 3443 & 4361 (+26.7\%) & 5368 (+29.2\%) & 41 & 43 (+4.9\%) & 46 (+7.2\%)\\ 
\hline
  Example~3 (count) & 1504 & 1737 (+15.5\%) & 1911 (+11.6\%) & 24 & 25 (+4.2\%) & 31 (+25.0\%) \\
\hline
  Example~3 (presence) & 1780 & 2095 (+17.7\%) & 2414 (+18.9\%) & 26 & 28 (+7.7\%) & 30 (+7.7\%)\\
\hline
\end{tabular}
\caption{Performance of examples from Section~\ref{sec:examples}. All
  time in ms. The percentages in parentheses in the column
  \textbf{IFC} are overheads relative to \textbf{Base}. Similar
  numbers in the column \textbf{\sys} are \emph{additional} overheads,
  still relative to \textbf{Base}.}
\label{table:eap}
\end{table}

\textit{JavaScript execution time:} To measure the overheads of
executing policy handler code, we interacted with all four programs
manually by entering relevant data and performing clicks a fixed
number of times. For each of these configurations, we measured the
total time spent \emph{only in executing JavaScript}, including
scripts and policies loaded initially with the page and the scripts
and policies executed in response to events. The difference between
\textbf{IFC} and \textbf{Base} is the overhead of taint tracking,
while the difference between the \textbf{\sys} and \textbf{IFC} is the
overhead of evaluating policy handlers. Since we are only measuring
JavaScript execution time and there are no time-triggered handlers in
these examples, variability in the inter-event gap introduced by the
human actor does not affect the measurements.

The left half of Table~\ref{table:eap} shows our observations.
All numbers are averages of 5 runs and the standard deviations are
all below 7\%. %\TODO{Fill these numbers}
Taint-tracking (\textbf{IFC}) adds overheads ranging from 15.5\% to
26.7\% over \textbf{Base}. To this, policy handlers (\textbf{\sys})
adds overheads ranging from 1.9\% to 29.2\%. \textbf{\sys} overheads
are already modest, but we also note that this is also a very
challenging (conservative) experiment for {\sys}. The scripts in both
sub-examples of Example~3 do almost nothing. The scripts in Examples~1
and Example~2 are slightly longer, but are still much simpler than
real scripts. On real and longer scripts, the relative overheads of
evaluating the policy handlers is significantly lower as shown
later. Moreover, our baseline in this experiment does not include
other browser costs, such as the cost of page parsing and rendering,
and network delays. Compared to those, both \textbf{IFC} and
\textbf{\sys} overheads are negligible.
%
%% In fact, the total overheads induced by
%% both taint tracking (\textbf{IFC}) and policy handlers (\textbf{\sys})
%% are too small to be perceived by the browser's user: For any
%% \emph{single} interaction, the total overhead never exceeds
%% ... ms. \TODO{Fill in the previous number. If it is more than 100ms,
%%   then delete the previous two sentences.}

\textit{Page load time:} We separately measured the time taken for
loading the initial page (up to the DOMContentLoaded event). The
difference between \textbf{\sys} and \textbf{IFC} is the overhead for
parsing and loading policies. The right half of Table~\ref{table:eap}
shows our observations.
All numbers are the average of 20 runs and all standard deviations are
below 8\%.
{\sys} overheads due to policy parsing and loading range from 7.2\% to
25\% (last column). When we add overheads due to taint tracking
(column \textbf{IFC}), the numbers increase to 12.1\% to 29.2\%. Note
that page-load overheads are incurred only once on every page (re-)load.

%%%%%%%%%%%%%%%%%%%%%%%%%%%%%%%%%%%%%%%%%%%%%%%%%%
%%% REAL-WORLD WEBSITES
%%%%%%%%%%%%%%%%%%%%%%%%%%%%%%%%%%%%%%%%%%%%%%%%%%

\medskip\noindent\textbf{Real-world websites.}
To understand whether {\sys} scales to real-world websites, we
evaluated {\sys} on policies for two real-world applications---the
website \url{ http://www.passwordmeter.com} that deploys a
password-strength checker (similar to Example~1) and a bank login page
that includes third-party analytics scripts (similar to
Example~3). Both policies were written by hand
and are shown in \iffull{Appendix~\ref{sec:realpolicies}}\ifconf{the
  full version of the paper~\cite{fullversion-esorics17}}.

\textit{Experience writing policies:} In both cases, we were able to
come up with meaningful policies easily after we understood the code,
suggesting that {\sys} policies can be (and should be) written by
website developers. The policy for the password-strength checker is
similar to Listing~\ref{egscript1} and prevents the password from
being leaked to third-parties. We had to write four lines of
additional policy code to allow the script to write the results of the
password strength check (which depends on the password) into the host
page.
The analytics script on the bank website communicates all
user-behavior to its server.  We specified a policy that disallows
exfiltration of keypresses on the username and the password text-boxes
to third-parties.

\begin{table}[tbp]
\centering
\begin{tabular}{ | C{2cm} || C{1.5cm} | C{1.5cm} | C{1.5cm} || C{1.5cm} | C{1.5cm} | C{1.5cm} |}
\hline
\multicolumn{1}{|c||}{ } &
\multicolumn{3}{ c ||}{JavaScript Execution Time} &
                                                       \multicolumn{3}{
                                                         c |}{Page Load Time} \\
\hline
 \multicolumn{1}{|c||}{Website} & \textbf{Base} & \textbf{IFC} & \textbf{\sys} & \textbf{Base} & \textbf{IFC} & \textbf{\sys} \\
\hhline{|=#=|=|=#=|=|=|} 
 \multicolumn{1}{|c||}{Password} & 79.5 & 115.5 (+45.3\%) & 126 (+13.2\%) & 303 & 429 (+41.6\%) & 441 (+4.0\%)\\  
\hline
 \multicolumn{1}{|c||}{Analytics} & 273.4 & 375.1 (+37.2\%) & 386.1 (+4.0\%) & 2151 & 2422 (+12.6\%) & 2499 (+3.6\%) \\ 
\hline
\end{tabular}
\caption{Performance on two real-world websites. All time in ms. The
  percentages in parentheses in the column \textbf{IFC} are overheads
  relative to \textbf{Base}. Similar numbers in the column
  \textbf{\sys} are \emph{additional} overheads, still relative to
  \textbf{Base}.}
\label{table:rap}
\end{table}

\textit{Performance overheads:} We also measured performance overheads
on the two websites, in the same configurations as for the synthetic
examples. Table~\ref{table:rap} shows the results. On real-world
websites, where actual computation is long, the overheads of {\sys}
are rather small. The overheads of executing policy handlers, relative
to \textbf{Base}'s JavaScript execution time, are 4.0\% and 13.2\%,
while the overheads of parsing and loading policies are no more than
4.0\%. Even the total overhead of \textbf{IFC} and \textbf{\sys} does
not adversely affect the user experience in any significant way.

% \textcolor{red}
{This experiment indicates that {\sys} is suitable for real-world
  websites.}

\section{Related Work}
\label{sec:related}

Browser security is a very widely-studied topic. Here, we describe
only closely related work on browser security policies and policy
enforcement techniques.

\medskip\noindent\textbf{Information flow control and script
  isolation.}  The work most closely related to our is that of Vanhoef
\emph{et al.}~\cite{csf14} on stateful declassification policies in
reactive systems, including web browsers. Their policies are similar
to ours, but there are significant differences. First, their policies
are attached to the browser and they are managed by the browser user
rather than website developers. Second, the policies have
coarse-granularity: They apply uniformly to all events of a certain
type. Hence, it is impossible to specify a policy that makes
keypresses in a password field secret, but makes other keypresses
public. Third, the enforcement is based on secure
multi-execution~\cite{SME}, which is, so far, not compatible with
shared state like the DOM.

%% Our work is most closely related to the work by
%% Vanhoef et al.~\cite{csf14} on stateful declassification policies for
%% event-driven programs, which was focussed on black-box secure
%% multi-execution. Their framework allowed policies to be attached to
%% different events, which would then be projected to the public level or
%% release some information via the release channel to the public
%% level. The main drawback of this approach is that policies are
%% extensional. For instance, one cannot specify a policy that only
%% keypresses in the password field are secret without making all
%% keypresses secret. Hence in the currency converter example, the policy
%% cannot distinguish keypresses in currency name field and in currency
%% value field making it difficult to be enforced.

COWL~\cite{cowl} enforces mandatory access control at
coarse-granularity. In COWL, third-party scripts are sandboxed. Each
script gets access to either remote servers or the host's DOM, but not
both. Scripts that need both must be re-factored to pass DOM elements
over a message-passing API (\texttt{postMessage}). This can be both
difficult and have high overhead. For scripts that do not need this
factorization, COWL is more efficient than solutions based on FGTT.

%% The COWL system~\cite{cowl} is another recent work in this area that
%% confines untrusted code using labels for mandatory access
%% control. However, COWL works at a coarse-grained level, i.e., it
%% provides protection at the level of iframes or compartments and by
%% using worker threads. The major limitation with COWL is the access of
%% public data by third-party scripts, which would still require an
%% explicit message passing from the host page. For instance, consider an
%% advertising script that displays advertisements based on certain
%% keywords appearing on the page that contains some sensitive
%% information like user name, contacts etc. With COWL, however, as the
%% advertising script can access the sensitive data also, it would need
%% to be confined. Hence, the host page would have to provide the
%% required data to the script via \texttt{postMessage}, which in turn
%% requires that the keywords be communicated to the host page by the
%% script before the host page runs the script. As the criterion of the
%% advertising script can change over time, it would need to inform the
%% host page, which in turn would require changes (possibly frequent) to
%% the script on the host page.

Mash-IF~\cite{mashif} uses static analysis to enforce IFC
policies. Mash-IF's model is different from {\sys}'s model. Mash-IF
policies are attached only to DOM nodes and there is no support for
adding policies to new objects or events. Also, in Mash-IF, the
browser user (not the website developer) decides what
declassifications are allowed. Mash-IF is limited to a JavaScript
subset that excludes commonly used features such as \texttt{eval} and
dynamic property access.

JSand~\cite{jsand} uses server-side changes to the host page to
introduce wrappers around sensitive objects, in the style of object
capabilities~\cite{millerphd}. These wrappers mediate every access by
third-party scripts and can enforce rich access policies. Through
secure multi-execution, coarse-grained information flow policies are
also supported. However, as mentioned earlier, it is unclear how
secure multi-execution can be used with scripts that share state with
the host page.

%% Another very similar work is Mash-IF~\cite{mashif}. The policies are
%% defined in terms of DOM objects and the enforcement mechanism is a
%% static analysis for a subset of JavaScript (mostly static features
%% excluding \TT{eval} and dynamic property access) that treats the language
%% constructs outside this subset as blackboxes. A reference monitor has
%% exclusive access to the policies and is used to monitor the executions
%% by intercepting all DOM calls, and to derive declassification rules
%% based on the static analysis of the function and user preferences and
%% choices. However, there is no separate specification mechanism that
%% allows policies to depend on or relate to events. Thus, one cannot
%% specify the confidentiality of events in the browser. 

{\sys} policies are enforced using an underlying IFC
component. Although, in principle, any IFC technique such as
fine-grained taint
tracking~\cite{jang10CCS,jsflow,post14,chudnov-ccs}, coarse-grained
taint tracking~\cite{cowl} or secure multi-execution~\cite{SME} can be
used with {\sys}, to leverage the full expressiveness of {\sys}'s
finely-granular policies, a fine-grained IFC technique is
needed. JSFlow~\cite{csf12,jsflow} is a stand-alone implementation of
a JavaScript interpreter with fine-grained taint tracking. Many
seminal ideas for labeling and tracking flows in JavaScript owe their
lineage to JSFlow, but since JSFlow is written from scratch it has
very high overheads. Building on ideas introduced by Just \emph{et
  al.}~\cite{just11PLASTIC}, our own prior work~\cite{csf15,post14}
implements fine-grained IFC in an existing browser engine, WebKit, by
modifying the JavaScript interpreter. The overheads are significantly
lower than JSFlow, which is why chose to integrate {\sys} with our own
work. Both JSFlow and our work include formal proofs that the taint
tracking is complete, relative to the abstractions of a formal model.
Chudnov and Naumann~\cite{chudnov-ccs} present another approach to
fine-grained IFC for JavaScript. They rewrite source programs to add
shadow variables that hold labels and additional code that tracks
taints. This approach is inherently more portable than that of JSFlow
or our work, both of which are tied to specific, instrumented
browsers. However, it is unclear to us how this approach could be
extended with a policy framework like {\sys} that assigns
state-dependent labels at runtime.

%% Most of the other previous work has been dedicated on enhancing the
%% security with improved access control mechanisms. The common web
%% standards like  same origin policy~\cite{sop}, content security
%% policy~\cite{csp} and cross-origin resource sharing~\cite{cors}
%% restrict content access by third-party scripts by isolating them or
%% selectively allowing certain scripts and functionality to
%% run. However, they do not allow limited access to some of the
%% confidential data thereby mostly exposing sensitive information to
%% untrusted code.  

\medskip\noindent\textbf{Access control.}  The traditional browser
security model is based on restricting scripts' access to data, not on
tracking how scripts use data. However, no model based on access
control alone can simultaneously allow scripts access to data they
need for legitimate purposes and prevent them from leaking the data on
the side. Doing so is the goal of IFC and {\sys}. Nonetheless,
  we discuss some related work on access control in web
  browsers.

The standard same-origin policy (SOP) and content-security policy
(CSP) were described in Section~\ref{sec:intro}. An additional, common
access policy---cross-origin resource sharing
(CORS)~\cite{cors}---relaxes SOP to allow some cross-origin requests.

Conscript~\cite{conscript} allows the specification of fine-grained
access policies on individual scripts, limiting what actions every
script can perform. Similarly, AdJail~\cite{adjail} limits the
execution of third-party scripts to a shadow page and restricts
communication between the script and the host page.
Zhou and Evans~\cite{zhouESORICS11} take a dual approach, where
fine-grained access control rules are attached to DOM elements. The
rules specify which scripts can and cannot access individual
elements. Along similar lines, Dong \emph{et al.}~\cite{ccs13crypton}
present a technique to isolate sensitive data using authenticated
encryption. Their goal is to reduce the size of the trusted computing
base.
ADsafe~\cite{adsafe} and FBJS~\cite{fbjs} restrict third-party code to
subsets of JavaScript, and use static analysis to check for
illegitimate access. Caja~\cite{caja} uses object capabilities to
mediate all access by third-party scripts. WebJail~\cite{webjail}
supports least privilege integration of third-party scripts by
restricting script access based on high-level policies specified by
the developer.
All these techniques enforce only access policies and cannot control
what a script does with data it has been provided in good faith.

\section{Conclusion}
\label{sec:conclusion}

Third-party JavaScript often requires access to sensitive data to
provide meaningful functionality, but comes with the risk that the
data may be leaked on the side. Information flow control in web
browsers can solve this problem. Within this context, this paper
proposed {\sys}, a mechanism for labeling sensitive data, dynamically
and at fine-granularity. {\sys} uses JavaScript for policy
specification, which makes it developer-friendly, and re-uses the
browser's event handling logic for policy interpretation, which makes
it easy to implement and improves the likelihood of easy portability
across browsers and versions.  Our evaluation indicates that {\sys}
has low-to-moderate overhead, even including the cost of information
flow control and, hence, it can be used on websites today.

\medskip
\noindent \textbf{Acknowledgments.}  We thank several anonymous
reviewers for their excellent feedback.  This work was funded in part
by the Deutsche For\-schungs\-ge\-mein\-schaft (DFG) grant
``Information Flow Control for Browser Clients'' under the priority
program ``Reliably Secure Software Systems'' (RS$^3$).

%% The problem of releasing sensitive data to semi-trusted code and
%% ensuring that the code does not leak the data is the basis of all work
%% on information flow control. In web browsers, this problem manifests
%% in the context of third-party scripts, which provide a useful service,
%% but may leak information under the hood. Building on a body of
%% literature on taint tracking in web browsers, this paper has proposed
%% {\sys}, a policy mechanism for labeling sensitive data.  {\sys} is
%% highly expressive and overloads familiar constructs (event handlers)
%% to ease deployment. {\sys} is designed to run over a fine-grained
%% taint tracking mechanism, which has moderate runtime overhead. It
%% could be interesting to understand whether this overhead can be
%% traded-off for some loss in expressiveness by integrating {\sys} with
%% lighter enforcement techniques, e.g., those based on coarse-grained
%% isolation.

% \section{Acknowledgments}
\ifconf{
\bibliographystyle{splncs03}
\bibliography{main-full}
}

\iffull{

\appendix
\section{Preventing overlay-based attacks in {\sys}}
\label{sec:overlay}

To bypass {\sys} policies, an adversarial script may ``trick'' a user
using transparent overlays. For example, suppose a script wants to
exfiltrate the contents of a password field that is correctly
protected by a {\sys} policy. The script can create a transparent
overlay on top of the password field. Any password the user enters
will go into the overlay, which \emph{isn't} protected by any policy
and, hence, the script can leak the password.

Such attacks can be prevented easily in {\sys} using a \emph{single}
policy, attached to the top element of the page, that labels data
entered into all significantly transparent overlays as
\texttt{HOST}. Listing~\ref{overlay} shows such a policy. This
particular policy labels all keypress events on elements of opacity
below $0.5$ as \texttt{HOST}, thus preventing their exfiltration. The
threshold value $0.5$ can be changed, and the policy can be easily
extended to other user events like mouse clicks.

\begin{lstlisting}[float, caption=Example policy to prevent overlay-based
  stealing of keystrokes,label=overlay,language=C]
document.body.addEventListener("keypress", function(event){
    var o = window.getComputedStyle(event.target).getPropertyValue("opacity");
    if (o < 0.5) 
         event.setLabel("HOST");
});
\end{lstlisting}

\section{User-study for {\sys}}
\label{sec:user-study}

To understand how easily programmers can use {\sys}, we conducted a
small study with six students from our university, recruited through
an open call.  All participants knew how to program as we specifically
asked for this skill in the call, but only four knew JavaScript well
and of those four, two had studied information flow control
previously. Each participant's goal was to write policies for four
scenarios very similar to Examples~1,~2 and the two sub-examples of
Example~3. The participants were given a document of approximately
2,000 words that explained {\sys} and its API, and provided an
illustrative example. The participants were allowed to ask questions
about the documentation. Then, the participants were given the code of
the host pages and included third-party scripts for the four scenarios
and were asked to implement policies. They did this on our running
prototype and could see the consequences of their mistakes in
real-time and could debug their policies. The participants had up to
one hour to study the documentation and implement the policies.

Although the study size is quite small to state results with statistical
confidence, our observations tend to indicate that {\sys} can be used
by JavaScript programmers with a bit of training. All participants
were able to complete three of the four exercises. The two
participants who knew about information flow control were able to
complete all four exercises. The exercise that the remaining four
participants could not complete involved context labels and
\texttt{setContext()}. This is unsurprising, since context labels are
a difficult concept, although they are not needed very often. The two
participants who did not know JavaScript well encountered another
policy independent hurdle: They did not know the syntax for writing
JavaScript handlers. Once they were explained this syntax, they were
able to write all policies except the one involving
\texttt{setContext()}.

Overall, this \emph{suggests} that JavaScript programmers should be
able to use {\sys} with little training, except the context
labels. With an understanding of how information flow tracking works,
they should be able to use context labels as well.

\section{{\sys} policies on real-world websites}
\label{sec:realpolicies}

The {\sys} policies we wrote for the password strength checking
website and the bank website with an analytics script are shown in
Listings~\ref{realpolicy1} and~\ref{realpolicy2}, respectively. The
code on lines~\ref{ex:start}--\ref{ex:end} of
Listing~\ref{realpolicy1} allows the strength-checking script to write
back the visual indicator of password strength to the host page's DOM.

% \TODO{Again, the previous paragraph uses cross-references to
%   lines. Make this consistent with the text.}

\begin{lstlisting}[float, caption=Policy code for password strength
  checking website,label=realpolicy1,language=C,escapechar=\%]
document.getElementById("passwordPwd").setLabel("secret");
document.getElementById("passwordTxt").setLabel("secret");
var x = document.getElementsByTagName("div"); %\label{ex:start}%
var i = 0;
for (i = 0; i < x.length; i++) 
    x[i].setLabel("secret"); %\label{ex:end}%
\end{lstlisting}

\begin{lstlisting}[float, caption=Policy code for bank login
  website with an analytics script,label=realpolicy2,language=C] 
var x = document.getElementsByClassName("user"); // username
var y = document.getElementsByClassName("pwd"); // password
for (i = 0; i < x.length; i++) {
    x[i].addEventListener("keypress", function(event){
    event.setLabel("HOST");
    });}
for (i = 0; i < y.length; i++) {
    y[i].addEventListener("keypress", function(event){
    event.setLabel("HOST");
    });}
\end{lstlisting}

}

\end{document}